\documentclass[11pt,english]{article}
\usepackage[T1]{fontenc}
\usepackage[latin9]{inputenc}
\usepackage{listings}
\usepackage{geometry}
\usepackage{amssymb}
\usepackage{graphicx}
\usepackage{babel}
\begin{document}

\title{A Matrix Public Key Cryptosystem}

\author{M. Andrecut}

\maketitle

{

\centering Calgary, AB, Canada

\centering mircea.andrecut@gmail.com

}

\begin{abstract}
We discuss a matrix public key cryptosystem and its numerical implementation.
\end{abstract}

\section{Introduction}

In a symmetrical key cryptosystem, such as AES (Advanced Encryption
Standard), two users Alice and Bob must first agree on a common secret
key {[}1{]}. If Alice communicates the secret key to Bob, a third
party, Eve, might intercept the key, and decrypt the messages. In
order to avoid such a situation one can use an asymmetric public key
cryptosystem, which provides a mechanism to securely exchange information
via open networks {[}1{]}. 

In public key cryptography a user has a pair of cryptographic keys,
consisting on a widely distributed public key and a secret private
key. These keys are related through a hard mathematical inversion
problem, such that the private key cannot be practically derived from
the public key. The two main directions of public key cryptography
are the public key encryption and the digital signatures. Public key
encryption is used to ensure confidentiality. In this case, the data
encrypted with the public key can only be decrypted with the corresponding
private key. Digital signatures are used to ensure authenticity or
prevent repudiation. In this case, any message signed with a user's
private key can be verified by anyone who has access to the user's
public key, proving the authenticity of the message. 

A standard implementation of public key cryptography is based on the
Diffie-Hellman (DH) key agreement protocol {[}2{]}. The protocol allows
two users to exchange a secret key over an insecure communication
channel. The platform of the DH protocol is the multiplicative group
$\mathbb{Z}_{p}$ of integers modulo a prime $p$. The DH protocol
can be described as following: 
\begin{enumerate}
\item Alice and Bob agree upon the public integer $g\in\mathbb{Z}_{p}$.
\item Alice chooses the secret integer $a$.
\item Alice computes $A=g^{a}\,\mathrm{mod}\, p$, and publishes $A$.
\item Bob chooses the secret integer $b$. 
\item Bob computes $B=g^{b}\mathrm{\, mod}\, p$, and publishes $B$.
\item Alice computes the secret integer $K_{A}=B^{a}\mathrm{\, mod\,}p=g^{ba}\,\mathrm{mod\,}p$.
\item Bob computes the secret integer $K_{B}=A^{b}\mathrm{\, mod\,}p=g^{ab}\mathrm{\, mod\,}p$.
\end{enumerate}
It is obvious that both Alice and Bob calculate the same integer $K\equiv K_{A}=K_{B}$,
which then can be used as a secret shared key for symmetric encryption.

Assuming that the eavesdropper Eve knows $p,g,A$ and $B$, she needs
to compute the secret key $K$, that is to solve the discrete logarithm
problem: 
\begin{equation}
A=g^{a}\mathrm{\, mod}\, p,
\end{equation}
for the unknown $a$. If $p$ is a very large prime of at least 300
digits, and $a$ and $b$ are at least 100 digits long, then the problem
becomes computationally hard (exponential time in $\log\, p$), and
it is considered infeasible. For maximum security it is also recommended
that $p$ is a safeprime, i.e. $(p-1)/2$ is also a prime, and $g$
is a primitive root of $p$. This protocol is secure, assuming that
it is infeasible to factor a large integer composed of two or more
large prime factors. 

Here, we discuss a public key cryptosystem, which extends the key
exchange to a matrix protocol, and avoids the traditional number theory
approach based on the discrete logarithm problem. The key exchange
mechanism is based on a commutative function defined as the product
of two matrix polynomials. We provide a detailed description of this
encoding mechanism, and a practical numerical implementation.

\section{Commutativity of Matrix Polynomials}

Let us consider the complex matrix $X\in\mathbb{C}^{N\times N},$
and two complex matrix polynomials of the form:
\begin{equation}
P(X,a):\mathbb{C}^{N\times N}\times\mathbb{C}^{M}\rightarrow\mathbb{C}^{N\times N},
\end{equation}
\begin{equation}
P(X,a)=\sum_{m=1}^{M}a_{m}X^{m},
\end{equation}
and
\begin{equation}
P(X,b):\mathbb{C}^{N\times N}\times\mathbb{C}^{K}\rightarrow\mathbb{C}^{N\times N},
\end{equation}
\begin{equation}
P(X,b)=\sum_{k=1}^{K}b_{k}X^{K},
\end{equation}
where $a\in\mathbb{C}^{M}$ and $b\in\mathbb{C}^{K}$. 

One can easily show that the product of these polynomials is commutative:
\[
P(X,a)P(X,b)=\left(\sum_{m=1}^{M}a_{m}X^{m}\right)\left(\sum_{k=1}^{K}b_{k}X^{k}\right)=
\]
\[
=\sum_{m=1}^{M}\sum_{k=1}^{K}a_{m}b_{k}X^{m+k}=
\]
\begin{equation}
=\left(\sum_{k=1}^{K}b_{k}X^{k}\right)\left(\sum_{m=1}^{M}a_{m}X^{m}\right)=P(X,b)P(X,a).
\end{equation}
However, because the matrix product is non-commutative, we also have:
\begin{equation}
P(X,a)P(Y,b)\neq P(Y,b)P(X,a),
\end{equation}
if $X\neq Y\in\mathbb{C}^{N\times N}$ and $a\in\mathbb{C}^{M}$,
$b\in\mathbb{C}^{K}$.

\section{Key Exchange Protocol}

Using the above property, the proposed key exchange protocol can be
formulated as following:
\begin{enumerate}
\item Alice chooses the secret vectors $a\in\mathbb{C}^{M_{1}}$ and $\tilde{a}\in\mathbb{C}^{M_{2}}$
(Alice's private key). 
\item Alice randomly generates and publishes the matrix $U\in\mathbb{C}^{N\times N}$
(Alice's matrix public key). 
\item Bob chooses the secret vectors $b\in\mathbb{C}^{J_{1}}$ and $\tilde{b}\in\mathbb{C}^{J_{2}}$
(Bob's private key). 
\item Bob randomly generates and publishes the matrix $V\in\mathbb{C}^{N\times N}$
(Bob's matrix public key). 
\item Alice computes and publishes the matrix $A=P(U,a)P(V,\tilde{a})$
(Alice's public key).
\item Bob computes and publishes the matrix $B=P(U,b)P(V,\tilde{b})$ (Bob's
public key).
\item Alice calculates the secret matrix $S_{a}=P(U,a)BP(V,\tilde{a})$
(Alice's secret key).
\item Bob calculates the secret matrix $S_{b}=P(U,b)AP(V,\tilde{b})$ (Bob's
secret key).
\end{enumerate}
One can see that both Alice and Bob obtain the same secret key $S\equiv S_{a}=S_{b}$,
since the matrix polynomials satisfy the commutativity property. Also
we assume that we always have $U\neq V$.

Assuming that the eavesdropper Eve knows $U,V$ and $A,B$, the hard
problem is to compute $S$, which requires the unknown private key
vectors $a,\tilde{a}$ or $b,\tilde{b}$. We should note that the
number of elements ($M_{1}$, $M_{2}$, $J_{1}$, $J_{2}$) in these
vectors is also unknown. 

One may attempt to find these quantities directly from $A$ and $B$.
For example we have:
\begin{equation}
A=\sum_{m=1}^{M_{1}}\sum_{n=1}^{M_{2}}a_{m}\tilde{a}_{n}U^{m}V^{n}=\sum_{k=1}^{K}\alpha_{k}\Psi^{(k)}.
\end{equation}
Thus, we obtain a system of equations,$\Psi\alpha=A$, with the unknown
$\alpha_{k}=a_{m}\tilde{a}_{n}$, $k\equiv(m,n)$. The columns of
the matrix $\Psi$, $\Psi^{(k)}=U^{m}V^{n}$, and the matrix $A$
are written as vectors by stacking their constituent columns. In total
there are $K=M_{1}M_{2}$ columns in $\Psi$, and each column has
$N^{2}$ elements. Therefore, the resulted system is badly underdermined
if $M_{1},M_{2}\gg N$. Thus, we should consider that $M_{1},M_{2},J_{1},J_{2}\gg N$,
in order to increase the security.

\section{Practical Implementation Aspects}

The goal of this section is to discuss some practical aspects regarding
the numerical implementation of the proposed matrix public key cryptosystem.
We choose the \texttt{C} language for implementation, and the standard
\texttt{GCC} compiler on a desktop running \texttt{Ubuntu 14.04 LTS}.
We focus only on building a simulation environment, where we can test
the functionality of the algorithms. The code is listed in the Appendix
1, and here we provide a detailed discussion of its main components.
The required compilation and run steps are: 
\begin{lstlisting}[basicstyle={\ttfamily},tabsize=2]
	gcc -O3 -lm keys.c -o keys
	./keys > results.txt
\end{lstlisting}

The program starts by setting the size of the required square matrices.
We choose the size value \texttt{N = 4}, which means that all the
matrices will contain \texttt{NN = N{*}N = 16} double precision complex
numbers. This value of \texttt{N} is enough to obtain a secret key
with the length \texttt{K = NN{*}8 = 128 bytes}. 

Since this is a simulation only, we choose also to use the standard
random number generator from \texttt{C}, \texttt{rand()}, in order
to initialize the variables in the problem. Of course in a real application
\texttt{rand()} is a poor choice, and should be avoided, however our
goal here is to simply show that the proposed algorithm works. For
a secure implementation one can use successive applications of secure
hash functions (\texttt{sha256)}, or secure ciphers (\texttt{AES}),
to generate the required pseudo-random numbers. 

The private keys, $a$, $\tilde{a}$, $b$, $\tilde{b}$, are stored
in the byte arrays \texttt{a}, \texttt{aa},\textsf{ }\texttt{b}, \texttt{bb},
and they are generated with the \texttt{private\_key(...)} function.
The private keys are stored as byte arrays of length \texttt{M1{*}16},
\texttt{M2{*}16},\textsf{ }\texttt{J1{*}16}, \texttt{J2{*}16}. Longer
private keys will provide a better security. For simulation purposes,
the \texttt{M1}, \texttt{M2}, \texttt{J1}, \texttt{J2} values are
randomly generated in the range \texttt{{[}NN, 2{*}NN-1{]}}, such
that the condition $M_{1},M_{2},J_{1},J_{2}\gg N$ is marginally satisfied
(one can make them much longer if necessary). 

Each double precision complex number needs \texttt{16 bytes} to store
it. The real and imaginary part of the complex numbers are generated
with the \texttt{drand()} function. This function is designed to generate
``dense'' double precision random numbers in the \texttt{(0, 1)}
interval. For each decimal in the mantissa, we use a separate \texttt{rand()}
call, until the precision gets below $2\cdot10^{-16}$, which is the
machine epsilon for double precision numbers. 

The matrix public keys, $U$ and $V$, are stored in the byte arrays
\texttt{U} and\textsf{ }\texttt{V}, and are generated with the \texttt{matrix\_public\_key(...)}
function. Each array holds \texttt{NN} double precision complex numbers
in \texttt{NN{*}16 bytes}, using the same approach described above
for the \texttt{private\_key(...)} function. 

The public keys, $A$ and $B$, are stored in the byte arrays \texttt{A}
and \texttt{B}, and are generated with the \texttt{public\_key(...)}
function. Since these are also double precision complex matrices with
the size \texttt{NN}, they will need \texttt{NN{*}16 bytes} of storage. 

The power matrices $U^{m}$ are also normalized as following:
\begin{equation}
\left\Vert U^{m}\right\Vert _{F}^{-1}U^{m}\leftarrow U^{m},
\end{equation}
where $\left\Vert .\right\Vert _{F}$ is the Frobenius norm. Thus,
all the polynomials are rewritten as:
\begin{equation}
P(U,a)=\sum_{m=1}^{M}a_{m}\left\Vert U^{m}\right\Vert _{F}^{-1}U^{m},
\end{equation}
This way, the elements of the power matrices will be in the same ``range'',
while the polynomial commuting properties described before are preserved.
Also, since the order $U$, $V$ or $V$, $U$ cannot be specified
exactly, we made the convention that if the first element of $U$
is smaller than the first element of $V$, then the order $U$, $V$
is switched to $V$, $U$. This makes irrelevant the order in which
the arguments, $U$ and $V$, are passed to the functions.

The secret keys, $S_{a}$ and $S_{b}$, are stored in the byte arrays
\texttt{Sa} and \texttt{Sb} with a length \texttt{K = NN{*}4}, and
they are computed with the \texttt{secret\_key(...)} function. The
secret keys are extracted from the matrices $S_{a}$ and $S_{b}$,
each containing \texttt{NN} elements. The computation of these matrices
is affected by rounding errors, due to the finite floating point number
representation, and inherently will lead to $S_{a}\neq S_{b}$, which
of course is undesirable. In order to counter the accumulation of
the floating point rounding errors, we extract the significand of
each double precision number, and from the significand we extract
a\texttt{ 4 byte} unsigned integer (see the code for more details).
Thus, from each double precision complex number we extract two unsigned
integers in the range $[0,2^{32}-1]$. The \texttt{128 bytes} corresponding
to these \texttt{32} integers form the secret keys, and are stored
into the byte arrays \texttt{Sa} and \texttt{Sb}. 

For illustration purposes, the results obtained for one instance run
are given in Appendix 2. One can see that both Alice and Bob compute
the same secret key (\texttt{Sa = Sb}), using completely different
secret and public keys.

\section{Conclusion}

In conclusion, we have presented a matrix public key cryptosystem,
which avoids the cumbersome key generation of the traditional number
theory approach. The key exchange mechanism is ensured by the commutative
property of the product of two matrix polynomials. Also, we have provided
a detailed description of this encoding mechanism, and a practical
numerical implementation.

\section*{Appendix 1}

Below we give the C code of the proposed key exchange protocol. 

\begin{lstlisting}[basicstyle={\small\ttfamily},breaklines=true,tabsize=2]
//------------------------------------------------------
// keys.c, C implementation of the key exchange protocol
// (c) 2015, M. Andrecut
//------------------------------------------------------

#include <stdlib.h>
#include <stdio.h>
#include <math.h>
#include <complex.h>
#include <time.h>

double drand(){
	double d = 0.0;
	double s = 1.0;
	do{
		s /= RAND_MAX + 1.0;
		d += rand() * s;
	}while(s > 2E-16);
	return d;
}

void cprod(unsigned N, double complex *X, 
		double complex *Y, double complex *R){
	unsigned i, j, n, NN = N*N;
	for(j=0; j<N; j++){
		for(i=0; i<N; i++){
			R[j*N+i] = 0;
			for(n=0; n<N; n++){
				R[j*N+i] += X[n*N+i]*Y[j*N+n];
			}
		}
	}
}


double fnorm(unsigned N, double complex *x){
	double norm = 0, d;
	unsigned n;
	for(n=0; n<N; n++){
		d = cabs(x[n]);
		norm += d*d;
	}
	return sqrt(norm);
}

void secret_key(unsigned M1, unsigned M2, 
				unsigned N, unsigned K, 
				unsigned char *a, unsigned char *aa, 
				unsigned char *U, unsigned char *V, 
				unsigned char *B, unsigned char *S){
	unsigned m, n, i, j, NN = N*N;
	double complex *u = calloc(NN, sizeof(double complex));
	double complex *v = calloc(NN, sizeof(double complex));
	double complex *pu = malloc(NN*sizeof(double complex));
	double complex *pv = malloc(NN*sizeof(double complex));
	if(U[0] >= V[0]){
		for(n=0; n<NN; n++){
			unsigned char* x = malloc(16);
			for(m=0; m<16; m++) x[m] = U[n*16+m];
			pu[n] = u[n] = *(double complex*)x;
		}
		for(n=0; n<NN; n++){
			unsigned char* x = malloc(16);
			for(m=0; m<16; m++) x[m] = V[n*16+m];
			pv[n] = v[n] = *(double complex*)x;
		}
	}
	else{
		for(n=0; n<NN; n++){
			unsigned char* x = malloc(16);
			for(m=0; m<16; m++) x[m] = V[n*16+m];
			pu[n] = u[n] = *(double complex*)x;
		}

		for(n=0; n<NN; n++){
			unsigned char* x = malloc(16);
			for(m=0; m<16; m++) x[m] = U[n*16+m];
			pv[n] = v[n] = *(double complex*)x;
		}
	}
	double complex *c = calloc(M1, sizeof(double complex));
	for(n=0; n<M1; n++){
		unsigned char* x = malloc(16);
		for(m=0; m<16; m++) x[m] = a[n*16+m];
		c[n] = *(double complex*)x;
	}
	double complex *cc = calloc(M2, sizeof(double complex));
	for(n=0; n<M2; n++){
		unsigned char* x = malloc(16);
		for(m=0; m<16; m++) x[m] = aa[n*16+m];
		cc[n] = *(double complex*)x;
	}
	double complex *uu = calloc(NN, sizeof(double complex));
	double complex *vv = calloc(NN, sizeof(double complex));
	double complex *r = calloc(NN, sizeof(double complex));
	for(m=0; m<M1; m++){
		double norm = fnorm(NN, pu);
		for(n=0; n<NN; n++) uu[n] += c[m]*pu[n]/norm;
		cprod(N, pu, u, r);
		for(n=0; n<NN; n++) pu[n] = r[n];
	}
	for(m=0; m<M2; m++){
		double norm = fnorm(NN, pv);
		for(n=0; n<NN; n++) vv[n] += cc[m]*pv[n]/norm;
		cprod(N, pv, v, r);
		for(n=0; n<NN; n++) pv[n] = r[n];
	}
	double complex *b = calloc(NN, sizeof(double complex));
	for(n=0; n<NN; n++){
		unsigned char* x = malloc(16);
		for(m=0; m<16; m++) x[m] = B[n*16+m];
		b[n] = *(double complex*)x;
	}
	cprod(N, uu, b, pu);
	cprod(N, pu, vv, r);
	for(n=0; n<NN; n++){
		double x = significand(creal(r[n]));
		if(x > 0) x = x - 1; else x = x + 2;
		unsigned fr = (unsigned)(x*4294967295);
		double y = significand(cimag(r[n]));
		if(y > 0) y = y - 1; else y = y + 2;
		unsigned fi = (unsigned)(y*4294967295);
		unsigned char* rr = (unsigned char*)&fr;
		unsigned char* ri = (unsigned char*)&fi;
		for(m=0; m<4; m++) S[n*8+m] = rr[m];
		for(m=0; m<4; m++) S[n*8+m+4] = ri[m];
		printf("%u\t%u\n", fr, fi);
	}
	free(u); free(v); free(c); free(cc);
	free(pu); free(pv); free(uu); free(vv);
	free(b); free(r);
}

void public_key(unsigned M1, unsigned M2, unsigned N, 
				unsigned char *a, unsigned char *aa, 
				unsigned char *U, unsigned char *V, 
				unsigned char *A){
	unsigned m, n, i, j, NN = N*N;
	double complex *u = calloc(NN, sizeof(double complex));
	double complex *v = calloc(NN, sizeof(double complex));
	double complex *pu = malloc(NN*sizeof(double complex));
	double complex *pv = malloc(NN*sizeof(double complex));
	if(U[0] >= V[0]){
		for(n=0; n<NN; n++){
			unsigned char* x = malloc(16);
			for(m=0; m<16; m++) x[m] = U[n*16+m];
			pu[n] = u[n] = *(double complex*)x;
		}
		for(n=0; n<NN; n++){
			unsigned char* x = malloc(16);
			for(m=0; m<16; m++) x[m] = V[n*16+m];
			pv[n] = v[n] = *(double complex*)x;
		}
	}
	else{
		for(n=0; n<NN; n++){
			unsigned char* x = malloc(16);
			for(m=0; m<16; m++) x[m] = V[n*16+m];
			pu[n] = u[n] = *(double complex*)x;
		}
		for(n=0; n<NN; n++){
			unsigned char* x = malloc(16);
			for(m=0; m<16; m++) x[m] = U[n*16+m];
			pv[n] = v[n] = *(double complex*)x;
		}
	}
	double complex *c = calloc(M1, sizeof(double complex));
	for(n=0; n<M1; n++){
		unsigned char* x = malloc(16);
		for(m=0; m<16; m++) x[m] = a[n*16+m];
		c[n] = *(double complex*)x;
	}
	double complex *cc = calloc(M2, sizeof(double complex));
	for(n=0; n<M2; n++){
		unsigned char* x = malloc(16);
		for(m=0; m<16; m++) x[m] = aa[n*16+m];
		cc[n] = *(double complex*)x;
	}
	double complex *uu = calloc(NN, sizeof(double complex));
	double complex *vv = calloc(NN, sizeof(double complex));
	double complex *r = calloc(NN, sizeof(double complex));
	for(m=0; m<M1; m++){
		double norm = fnorm(NN, pu);
		for(n=0; n<NN; n++) uu[n] += c[m]*pu[n]/norm;
		cprod(N, pu, u, r);
		for(n=0; n<NN; n++) pu[n] = r[n];
	}
	for(m=0; m<M2; m++){
		double norm = fnorm(NN, pv);
		for(n=0; n<NN; n++) vv[n] += cc[m]*pv[n]/norm;
		cprod(N, pv, v, r);
		for(n=0; n<NN; n++) pv[n] = r[n];
	}
	cprod(N, uu, vv, r);

	for(n=0; n<NN; n++){
		double complex cr = r[n];
		unsigned char* rr = (unsigned char*)&cr;
		for(m=0; m<16; m++) A[n*16+m] = rr[m];
		printf("%+.16lf\t%+.16lf\n", creal(cr), cimag(cr));
	}
	free(u); free(v);
	free(c); free(cc);
	free(pu); free(pv);
	free(uu); free(vv); free(r);
}

void matrix_public_key(unsigned N, unsigned char *U){
	unsigned n, m;
	for(n=0; n<N/16; n++){
		double complex c = drand()*2 - 1 + I*(drand()*2 - 1);
		unsigned char* array = (unsigned char*)&c;
		for(m=0; m<16; m++) U[n*16+m] = array[m];
		printf("%+.16lf\t%+.16lf\n", creal(c), cimag(c));
	}
}

void private_key(unsigned N, unsigned char *a){
	unsigned m, n;
	for(n=0; n<N/16; n++){
		double complex c = drand()*2 - 1 + I*(drand()*2 - 1);
		unsigned char* array = (unsigned char*)&c;
		for(m=0; m<16; m++) a[n*16+m] = array[m];
		printf("%+.16lf\t%+.16lf\n", creal(c), cimag(c));
	}
}

void sprint(unsigned N, unsigned char *x){
	unsigned n;
	for(n=0; n<N; n++){
		if(n % 32 == 0)
			printf("\n%02x", x[n]);
		else 
			printf("%02x", x[n]);
	}
	printf("\n");
}

unsigned char check_key(unsigned N, unsigned char *x, unsigned char *y){
	unsigned n, c = 0;
	for(n=0; n<N; n++){
		if(x[n] != y[n]){
			c = 1; 
			break;
		}
	}
	return c;
}


main(int argc, char *argv[]){

	// Dimension of matrices, NxN
	unsigned N = 4, NN = N*N;

	printf("Dimension of matrices\n");
	printf("---------------------\n");
	printf("N = %u\tNN = %u\n", N, NN);

	// Length of the secret key
	unsigned K = NN*8;
	printf("\nLength of the secret key\n");
	printf("------------------------\n");
	printf("K = %u\n", K);

	// Initialize the random number generator
	srand((unsigned char) time(0) + getpid());

	// Alice: private key vectors
	unsigned M1 = NN + rand()%NN, M2 = NN + rand()%NN;
	unsigned char *a = malloc(M1*16);
	unsigned char *aa = malloc(M2*16);
	printf("\nAlice: private key a\n");
	printf("--------------------\n");
	printf("\nM1 = %d complex numbers\n", M1);
	private_key(M1*16, a);
	printf("\nHex representation of a, %u bytes", M1*16);
	sprint(M1*16, a);
	printf("\nAlice: private key aa\n");
	printf("---------------------\n");
	printf("\nM2 = %d complex numbers\n", M2);
	private_key(M2*16, aa);
	printf("\nHex representation of aa, %u bytes", M2*16);
	sprint(M2*16, aa);

	// Bob: private key vectors
	unsigned J1 = NN + rand()%NN, J2 = NN + rand()%NN;
	unsigned char *b = malloc(J1*16);
	unsigned char *bb = malloc(J2*16);
	printf("\nBob: private key b\n");
	printf("------------------\n");
	printf("\nJ1 = %d complex numbers\n", J1);
	private_key(J1*16, b);
	printf("\nHex representation of b, %u bytes", J1*16);
	sprint(J1*16, b);
	printf("\nBob: private key bb\n");
	printf("-------------------\n");
	printf("\nJ2 = %d complex numbers\n", J2);
	private_key(J2*16, bb);
	printf("\nHex representation of bb, %u bytes", J2*16);
	sprint(J2*16, bb);


	// Alice: matrix public key
	unsigned char *U = malloc(NN*16);
	printf("\nAlice: matrix public key U\n");
	printf("--------------------------\n");
	printf("\nN*N = %d complex numbers\n", NN);
	matrix_public_key(NN*16, U);
	printf("\nHex representation of U, %u bytes", NN*16);
	sprint(NN*16, U);

	// Bob: matrix public key
	unsigned char *V = malloc(NN*16);
	printf("\nBob: matrix public key V\n");
	printf("------------------------\n");
	printf("\nN*N = %d complex numbers\n", NN);
	matrix_public_key(NN*16, V);
	printf("\nHex representation of V, %u bytes", NN*16);
	sprint(NN*16, V);

	// Alice: public key
	unsigned char *A = malloc(NN*16);
	printf("\nAlice: public key A\n");
	printf("-------------------\n");
	printf("\nN*N = %d complex numbers\n", NN);
	public_key(M1, M2, N, a, aa, U, V, A);
	printf("\nHex representation of A, %u bytes", NN*16);
	sprint(NN*16, A);

	// Bob: public key
	unsigned char *B = malloc(NN*16);
	printf("\nBob: public key B\n");
	printf("-----------------\n");
	printf("\nN*N = %d complex numbers\n", NN);
	public_key(J1, J2, N, b, bb, U, V, B);
	printf("\nHex representation of B, %u bytes", NN*16);
	sprint(NN*16, B);

	// Alice: secret key
	unsigned char *Sa = malloc(K);
	printf("\nAlice: secret key Sa\n");
	printf("--------------------\n");
	printf("\n2*N*N = %d unsigned integers on 4 bytes\n", 2*NN);
	secret_key(M1, M2, N, K, a, aa, U, V, B, Sa);
	printf("\nHex representation of Sa, %u bytes", K);
	sprint(K, Sa);

	// Bob: secret key
	unsigned char *Sb = malloc(K);
	printf("\nBob: secret key Sb\n");
	printf("------------------\n");
	printf("\n2*N*N = %d unsigned integers on 4 bytes\n", 2*NN);
	secret_key(J1, J2, N, K, b, bb, U, V, A, Sb);
	printf("\nHex representation of Sb, %u bytes", K);
	sprint(K, Sb);

	// Check the secret keys
	printf("\nCheck the secret keys\n");
	printf("---------------------\n");
	if(check_key(K, Sa, Sb) == 0){
		printf("Result: Sa = Sb\n");
	}
	else{
		printf("Result: Sa != Sb\n");
	}
	return 0;
}
\end{lstlisting}

\section*{Appendix 2}

For illustration purposes, the results obtained for one instance run
are given below. 

\begin{lstlisting}[basicstyle={\small\ttfamily}]

Size of matrices
----------------
N = 4	NN = 16

Length of the secret key
------------------------
K = 128

Alice: private key a
--------------------

M1 = 24 complex numbers
+0.8689850983830614	-0.8548606934416402
-0.1840404501840445	-0.9778283575338482
-0.5624424497754633	-0.3926427336457170
+0.2681173127751544	-0.2169790138199983
+0.6868307886138416	-0.9872768015372509
-0.5409692556398616	-0.8628458723853144
+0.0340424836674749	-0.3506396088873717
+0.4317577723105932	+0.7887667341912554
+0.0672014622485229	+0.2782204334161911
+0.9595978984392544	-0.0538938046045783
+0.7195946921855143	+0.8729643758373127
+0.5863311031238485	-0.7404569811244905
+0.7634357203607813	+0.6367243534294711
+0.2165438627036125	+0.5814044504531037
+0.2439741527589443	+0.7957204292763893
+0.5183932032859373	-0.9873300964591410
-0.4087175890232525	+0.9764747939005469
-0.5239555517482943	-0.5312443242637066
-0.0022945854777570	-0.0519642428515297
+0.4798449924309158	+0.1833511453978887
-0.4060460210598060	+0.8571984084524591
+0.6693677004712948	+0.4657471765812942
-0.1540452316059945	-0.9389158129164289
+0.3646444266501574	+0.5890737755058091

Hex representation of a, 384 bytes
ea8748d6b9ceeb3f9df71ed0045bebbf483e5731a38ec7bf5eb216b25e4aefbf
1b5ef54e87ffe1bf204e01fd0e21d9bf90157784d528d13f903a23e4f7c5cbbf
526fdf8f84fae53f9576d684c597efbf5faba3c19e4fe1bfc1456cf26e9cebbf
c0860d34046ea13f9e9f361de170d6bf3c8af759eba1db3ffac5f0bb933de93f
00cd99721d34b13fc07ea6135dced13ff26be3a606b5ee3fa0a47700f897abbf
aafba972eb06e73ffe6a99fc52efeb3fd06c117239c3e23f54c9c0d6d2b1e7bf
3a8771bf106ee83f7c7b51c00b60e43f68133b94b5b7cb3fe43c8e81dd9ae23f
f0a095878b3acf3f9c0990b08a76e93ff29dd257ad96e03fcfc054493598efbf
dee889d16d28dabf627a2511483fef3f4221ea6e3ec4e0bf60bedc18f4ffe0bf
0008b73218cc62bf60aa37a70e9baabf3cf868c5c7b5de3f688395e20c78c73f
dcdb4773a8fcd9bfea914f5b2b6eeb3fe2bbd0cf756be53f64bee73ecdcedd3f
741eed0fc1b7c3bfcf8bc52c990beebfe463c8935556d73f7a51173fb1d9e23f

Alice: private key aa
---------------------

M2 = 22 complex numbers
-0.5118704412976753	-0.0500573637169195
+0.7604183635929800	-0.4931928538884157
-0.3375783216472882	+0.1598131473912503
+0.1632397906851764	+0.3421602938986694
-0.3620384387433089	+0.3855473826440021
-0.8872378358483427	+0.5676916094065314
+0.7025298861832101	+0.2206711130955801
-0.6247065410609323	+0.3230649032942832
+0.6917058497340922	+0.0231011495289499
-0.0152488020307338	+0.8275970245954227
+0.3305744647758120	-0.7897520094890274
-0.2482330123079071	+0.7894395461742834
-0.9737185128877407	-0.7489919758449641
+0.4474754300516699	+0.6970578097454476
+0.7274738114424169	+0.0977605819469343
-0.3061669770440641	-0.7828203718630750
-0.2298109276249133	-0.0338097119771229
+0.7811974836736504	+0.8376630850857776
+0.6003193600063923	+0.4021131457350087
-0.6334603792724227	+0.6055862039104507
+0.6752938616270538	+0.7418148947846244
-0.6433983716975780	-0.4681352436902144

Hex representation of aa, 352 bytes
0c34a51e3e61e0bf80212d681ea1a9bf261c5de45855e83ff29584c27890dfbf
3213d41ae29ad5bf6814c2d8c174c43f3009339d0ae5c43fcc35124af4e5d53f
c8089345a32bd7bfb8ebe0edceacd83f23c2179a4064ecbf48ab1398872ae23f
dcd2b3f41f7be63f786ef576f33ecc3f0c876e9298fde3bf2c99886a18add43f
04ea614e7422e63f404ae8e6d3a7973f80913390c33a8fbf9aef5cc1ac7bea3f
50b5c6cc2128d53fb8949601a645e9bf1034d36e19c6cfbfa89a1fb91643e93f
6d9a0bbab328efbfb00e2705bef7e7bf247875fc6fa3dc3f18e6082e4c4ee63f
ceee9a287747e73f80cd4c66d606b93f7a3e61603d98d3bf436cf94edd0ce9bf
5cca34c9716acdbf503575ae814fa1bffa1083dd91ffe83ff05070d022ceea3f
66424cf2d035e33f8c4e8ec638bcd93f902d89b34e45e4bf96889651f660e33f
7e165cdf019ce53ffc231997f2bce73f38b4972eb896e4bf18497086edf5ddbf

Bob: private key b
------------------

J1 = 31 complex numbers
-0.8131572898736764	-0.9495491783647080
+0.1298227154186107	+0.7302350146477263
-0.7446356426328338	+0.9468203098361780
-0.1602102571093867	+0.0153968761011387
-0.0321974180566288	-0.1509698059198559
+0.6083909207849130	-0.6980298260647722
+0.1000333272410052	-0.8515098187419327
-0.4108327516259943	+0.7698810134887732
-0.4353843212160295	+0.8627621712051379
-0.9065787271243225	+0.9290429557182074
+0.6038810599955362	-0.3380758062243528
-0.7069626958783282	-0.9937368694489492
-0.9291368614924062	-0.1109532512441092
+0.1682809573220776	+0.9704573627930386
-0.3759606473197272	+0.8195097335932378
+0.4097225812747536	+0.6411222197176920
+0.0968815767756941	+0.9892165586232851
+0.7941227780332327	+0.6687766533299828
-0.9919780830175842	+0.5390884771002475
-0.5993550677175068	+0.8416746390522734
+0.9217383104992827	+0.5618991062378083
+0.6513040121506304	-0.8492378708487287
+0.5596626098265201	+0.3583335523844433
-0.4966598369750350	+0.4885499630912138
+0.3960260341933464	-0.0562765497713249
-0.0567990502670347	-0.2094754124106146
+0.8749646723409310	-0.4118473497446520
+0.1605486771157856	+0.2522820859568238
-0.6534623397058436	-0.2528823245864887
-0.1318483317593858	-0.8276937454948939
-0.4699044210083154	-0.7268149655182780

Hex representation of b, 496 bytes
a75dd06f6205eabf97a460f5b462eebfc81b80de079ec03f4acb49d2155ee73f
286891200ed4e7bf4aeb3d1b5a4cee3f6c60620bc581c4bf000785ba65888f3f
b04123132e7ca0bfd8fb8d85fa52c3bf7e8b7e3cf077e33fa09552a54256e6bf
609b02bdc89bb93f0d06f784913febbf5eff1674154bdabf16dcd781dda2e83f
ac18343356dddbbf26aab169bf9beb3fdbf10764b102edbfae71ec4ab8baed3f
3a596b5ffe52e33ffe5ad3b408a3d5bfe042493b709fe6bfe69a6343b1ccefbf
61c6333a7dbbedbf30737aa96e67bcbf58731efc3a8ac53fe6786b99fc0def3f
d88e343fbd0fd8bf102c117a6c39ea3f487fc00fe538da3f2ea5cdbe1284e43f
2026d6233bcdb83f8e5bfe7ba9a7ef3f342a152c7469e93fec2dcc4b9e66e53f
f6181dd248beefbfe678597a3640e13fdc3bd1adea2de3bf8e4413a7ffeeea3f
20126257e17eed3f302d9ed513fbe13ff018fe827bd7e43f203d3ae6f42cebbf
16f7bf8fc1e8e13f743e23daefeed63f8cd8425746c9dfbfd0147c106744df3f
68664e947d58d93f805f99a947d0acbff079aedec314adbf6852cf1e17d0cabf
4a829be9b5ffeb3f543e86fcb45bdabf306ed0eadb8cc43f2c4723c36325d03f
238946da29e9e4bf9e777558392fd0bffc2778f867e0c0bf5d230098777ceabf
787f1efee912debfb19e64751142e7bf

Bob: private key bb
-------------------

J2 = 17 complex numbers
+0.2395015183561122	+0.3971994682370852
-0.0465769789069043	-0.3663425200453365
+0.7537951322336816	+0.8256763657073602
-0.1563436965923156	-0.1521531633898803
-0.0856898852384244	+0.9043694082667142
+0.5493668891763306	+0.0864919672072155
-0.2105619723294515	+0.2586894465691962
+0.2936094795551767	-0.1546671663785663
-0.4165447796757208	-0.6470306531837624
-0.5891911056850502	+0.1686385513920701
+0.4081439302681709	+0.4335022113121134
-0.6185220295108234	+0.7933638301919215
-0.6507796759984934	-0.7439947848693523
-0.3987045968811376	-0.4433641387621932
+0.3471997888579212	-0.6838942312678632
+0.4039709032807668	-0.2897329644110010
-0.9501367908238625	+0.1739202672557096

Hex representation of bb, 272 bytes
b846575afca7ce3fa0478451b76bd93f30a34d12f0d8a7bf08a7aee52772d7bf
2c7c1af8161fe83f4a6279d7f06bea3ff45de6fb1103c4bf98095f3ec179c3bf
4870b2b6c5efb5bfe444001d98f0ec3f2690d0de6994e13fd08d856a5624b63f
14d777d8b1f3cabf4071352e5e8ed03f4c0b1f6a7fcad23f0c33ae3a22ccc3bf
36b6816faba8dabf82e1dda079b4e4bf255d404ea7dae2bfd87089b3f295c53f
400324b8071fda3fc011150f80bedb3f565813b6eecae3bf7e1d108b3c63e93f
86142ae62fd3e4bfcf11ad26cecee7bf6ca317496084d9bf18fb0cfb1360dcbf
04aa94768538d63f39fea62776e2e5bf201888c6a8dad93ff6f9ad21fc8ad2bf
84136a458567eebff8c7fcf10443c63f

Alice: matrix public key U
--------------------------

N*N = 16 complex numbers
+0.3368737329532423	+0.5243043640910983
+0.2391556223941222	-0.3747238699602812
-0.8836268003784012	-0.0031821031027317
-0.4046797065655433	-0.0835783218659490
+0.2883828453162156	-0.6358192650617027
+0.8417251050926819	-0.4126508745199150
-0.6391474565256796	+0.2763919025872350
-0.2071290143032435	+0.2331221310471723
+0.7152330705414824	-0.1286121438584548
+0.9153039464129042	-0.1252376190599107
+0.1244033889339511	-0.4249777151124192
-0.0354984985378332	-0.5027293312737142
+0.5417379064036945	-0.0836606229409183
-0.1986953363917494	-0.1115330181161580
-0.3848289466173010	-0.3708409810614971
-0.3802923558073766	+0.5728729606643514

Hex representation of U, 256 bytes
68997ad8568fd53fa67b1df219c7e03f002c6bc4a69cce3f42a3a0d379fbd7bf
bfd12fb6ab46ecbf008a47985a116abfda7243b645e6d9bfb89d118f6365b5bf
a0135752dd74d23fa66ab3a4a158e4bf240cd37c69efea3f3aa5ae36df68dabf
162ee35de573e4bff8729fa967b0d13f4c193e1b3483cabfa824692cf2d6cd3f
58c9df7630e3e63f0cc9dedb5c76c0bff8c777802b4aed3f30ad0b4bc907c0bf
909efb86e6d8bf3f00f2fbbad532dbbfb0df8df4db2ca2bf0cf091d25b16e0bf
6e3ce0bbea55e13ff0857e57c86ab5bfd0cbd549d96ec9bfc0aa3b896d8dbcbf
c6a0119709a1d8bf6e3e6bcfdbbbd7bf7622c7bfb556d8bf5c1ddaacf954e23f

Bob: matrix public key V
------------------------

N*N = 16 complex numbers
+0.9409248020558381	+0.5377867567981318
-0.7158705844244213	+0.2893294067383878
+0.3557956546476002	+0.8879058945032454
+0.2466279650314691	+0.1282564850152923
+0.5482843655448633	-0.4268184961805279
+0.6265073683117459	-0.8281711179997823
-0.3367309500069268	+0.5117462362356848
+0.5288800137059331	+0.6497204598425779
+0.6000563534476080	+0.6179174079884797
-0.2284014936471640	+0.3260335651618362
+0.5991160201006454	+0.2927185726013939
-0.5344290793776173	-0.5523086828897967
-0.5174786942983332	+0.6341098574986008
-0.3930733586311385	-0.6365556865422852
-0.0401902524752414	-0.6492929052234647
+0.1879347693971776	-0.4468276153836913

Hex representation of V, 256 bytes
34679a540e1cee3fac6f95928c35e13f76ae886d69e8e6bfe898ed7c5f84d23f
784d31235bc5d63f66255a9fb969ec3fa06a0b528191cf3f00ff5160b56ac03f
a68b5da78b8be13fd4169b86fe50dbbf6243332e590ce43f55a069b76080eabf
322b75f8ff8cd5bfa8778fa43960e03f1e004cc795ece03f4a1cd28f82cae43f
5410ba61a933e33f48115ebafac5e33f443bc998423ccdbf002df1e2bbddd43f
42f11a5cf52be33f744e10aee6bbd23f05aa3e030b1ae1bf54df494283ace1bf
806b8c7a2f8fe0bf02e380c1a04ae43fc2f90f291d28d9bfac05f907aa5ee4bf
f08fff17d193a4bf1db52eea01c7e4bf88cc2b1c3f0ec83f9e72c1dad298dcbf

Alice: public key A
-------------------

N*N = 16 complex numbers
-0.2856153551337328	-0.0836362385140290
+0.0053028357754545	-0.6553666223075714
-0.6848378625559008	+1.0033521606714080
-0.1177416781224439	+0.1881147013745262
-0.1356525197048316	-0.4999190536353746
-0.0119241534869403	+0.1657257251876224
-0.1496198348578027	-0.2369035535027847
+0.1844067261375284	+0.0291454067840425
+0.1145145558807253	+0.0391101989223119
-0.1005465316239941	-0.6368000194347039
-0.4039634042904244	+1.0076922004908939
-0.2481990593712110	+0.2515002673868378
+0.1218629445687005	+0.0144359211753338
+0.5246569994564118	+0.0595147923884747
-0.5467684445921881	-0.4633383063630160
-0.2094934354311639	-0.2169002141819963

Hex representation of A, 256 bytes
4c3a62a08547d2bf309f2d3d2f69b5bfa0c7b4236db8753f9e6e366cc3f8e4bf
34afd71731eae5bf2f46c7feba0df03f747ce9905124bebf6273267d2414c83f
a249e0cf0f5dc1bf90b7807cacfedfbf803612cab06b88bf4cb3e4248036c53f
1c0ac624be26c3bffee14c0bdb52cebf9128f6bca39ac73fc843dd234bd89d3f
42766c70d350bd3fc26d9f824006a43ff8200de16abdb9bf120e326faa60e4bf
22538d5289dad9bfc1ae58db811ff03f2adb729dfcc4cfbf7b26d7939418d03f
f34e84f16832bf3f857faf8a94908d3f850cc979fdc9e03fe09adf0db978ae3f
78438089207fe1bf2e0b34b655a7ddbf15a7f34eaed0cabfdb1a34df62c3cbbf

Bob: public key B
-----------------

N*N = 16 complex numbers
-2.4633752471535728	-0.1602678155173908
-1.7414343240690795	-1.0301803412988897
-0.7089011929436275	+2.6033653995146535
+0.2888098643832883	+1.2726176834487517
-0.4447799179419522	+0.2067805379384524
-0.1946598847601872	-0.0414232823029533
-0.0495878578361504	+0.4089135120393886
+0.0332765292089976	+0.0188205856998400
-2.0814149405134770	+0.5097364399723628
-1.6706358918192217	-0.2970409670111223
+0.2365166391262642	+2.1543529375391444
+0.6503887269428416	+0.8676760384430399
+0.6959317836726924	-0.6902765198764746
+0.7908533006696801	-0.2909750558709320
-0.6917520195247205	-0.9240611824254524
-0.5250685199630863	-0.1903208570713778

Hex representation of B, 256 bytes
7367e214feb403c0a6d01fe1a783c4bf1223e03ceadcfbbf0ac6ad619e7bf0bf
7a39f98d51afe6bf96a0133db1d304407275925edc7bd23f26942b5ca45cf43f
9e9b5e304677dcbf678df2dfc877ca3f57ae71779deac8bf707d91b56e35a5bf
80b35e679463a9bfde814694a32bda3f20a45a099f09a13f821d4a20b445933f
134a57e0bca600c0315868cbc24fe03f97326eb3ecbafabf99a3b81db802d3bf
ede9005f2d46ce3f602d96641d3c0140b102fd04fccfe43f9d57148a00c4eb3f
8ce063bb1245e63fec1dc2c8be16e6bff4eac994ab4ee93f1bb73ad7559fd2bf
bf9e9921d522e6bf2fa7c0c1e891edbf8bd12c7f5ccde0bf13236f106f5cc8bf

Alice: secret key Sa
--------------------

2*N*N = 32 unsigned integers on 4 bytes
3532271144	1241794686
872140306	1131329665
3218067163	1418852160
1175145558	3891638795
2871309341	18700978
380671065	746843244
1259174687	3031792119
3387166373	3348009390
1308192945	820672386
212774227	4220132680
4247529633	3762180882
4198654425	3427153554
2079979731	1375907524
3773268400	1991239246
1322895297	3617498003
4146361598	273775259

Hex representation of Sa, 128 bytes
282e8ad27e48044a12cefb3381b86e43dbcecfbf40f79154564c0b460bb2f5e7
1db424abb25a1d015994b0166cec832c1f7b0d4bf779b5b4a50ee4c9ae918ec7
b170f94d8277ea3053adae0c481d8afba1282cfd12533ee0d96142fa923646cc
d3f8f97bc4ae0252b081e7e04ee6af76c1c7d94e93a39ed7fe7424f79b7a5110

Bob: secret key Sb
------------------

2*N*N = 32 unsigned integers on 4 bytes
3532271144	1241794686
872140306	1131329665
3218067163	1418852160
1175145558	3891638795
2871309341	18700978
380671065	746843244
1259174687	3031792119
3387166373	3348009390
1308192945	820672386
212774227	4220132680
4247529633	3762180882
4198654425	3427153554
2079979731	1375907524
3773268400	1991239246
1322895297	3617498003
4146361598	273775259

Hex representation of Sb, 128 bytes
282e8ad27e48044a12cefb3381b86e43dbcecfbf40f79154564c0b460bb2f5e7
1db424abb25a1d015994b0166cec832c1f7b0d4bf779b5b4a50ee4c9ae918ec7
b170f94d8277ea3053adae0c481d8afba1282cfd12533ee0d96142fa923646cc
d3f8f97bc4ae0252b081e7e04ee6af76c1c7d94e93a39ed7fe7424f79b7a5110

Check the secret keys
---------------------
Result: Sa = Sb

\end{lstlisting}

\end{document}